\documentclass[onecolumn,nofootinbib,preprintnumbers,superscriptaddress,longbibliography,
10pt]{revtex4-1}

\usepackage{amssymb}
\usepackage{graphics}
\usepackage{graphicx}
\usepackage{amsmath}
\usepackage{amsfonts}
\usepackage{bm}

\newlength{\extraspace}
\newlength{\extraspaces}
\newcommand{\be}{\begin{equation}
\addtolength{\abovedisplayskip}{\extraspaces}
\addtolength{\belowdisplayskip}{\extraspaces}
\addtolength{\abovedisplayshortskip}{\extraspace}
\addtolength{\belowdisplayshortskip}{\extraspace}}
\newcommand{\ee}{\end{equation}}

\newcommand{\ba}{\begin{eqnarray}
\addtolength{\abovedisplayskip}{\extraspaces}
\addtolength{\belowdisplayskip}{\extraspaces}
\addtolength{\abovedisplayshortskip}{\extraspace}
\addtolength{\belowdisplayshortskip}{\extraspace}}
\newcommand{\ea}{\end{eqnarray}}

\usepackage{color,soul}
\usepackage[colorlinks,citecolor=blue,urlcolor=blue,linkcolor=blue]{hyperref}

\begin{document}

\title{Rotating AdS  black holes in Maxwell-$f(T)$ gravity}

 \author{G.G.L. Nashed}
\email{nashed@bue.edu.eg}
\affiliation {Centre for Theoretical Physics, The British University, P.O. Box
43, El Sherouk City, Cairo 11837, Egypt}
\affiliation {Mathematics Department, Faculty of Science, Ain Shams University, Cairo
11566, Egypt}

\author{Emmanuel N. Saridakis}
\email{Emmanuel\_Saridakis@baylor.edu}
\affiliation{Department of Physics, National Technical University of Athens, Zografou
Campus GR 157 73, Athens, Greece}
\affiliation{CASPER, Physics Department, Baylor University, Waco, TX 76798-7310, USA}

\begin{abstract}
The investigation on higher-dimensional AdS black holes is of great importance under the
light of  AdS/CFT correspondence. In this work we study static and rotating, uncharged
and charged, AdS black holes in higher-dimensional $f(T)$ gravity, focusing on the
power-law ansatz which is the most viable according to observations. We extract AdS
solutions  characterized by an effective cosmological constant that depends on the
parameters of the $f(T)$ modification, as well as on the electric charge, even if the
explicit cosmological constant is absent. These solutions do not have a general
relativity or an uncharged limit, hence they correspond to a novel solution class, whose
features arise solely from the torsional modification alongside the Maxwell sector
incorporation. We examine the singularities of the solutions,
calculating the values of various curvature and torsion invariants, finding that they do
possess the central singularity, which however  is softer comparing to   standard
general relativity case due to the $f(T)$ effect. Additionally, we investigate the
horizons structure, showing that the solutions possess an inner Cauchy horizon as well
as an outer event  one, nevertheless for suitably large electric charge and
small mass we obtain the appearance of a naked
singularity. Finally, we calculate the energy
of the obtained solutions, showing that the $f(T)$ modification affects the mass term.

\end{abstract}

\maketitle

\section{Introduction}\label{S1}

After the formulation of the AdS/CFT correspondence
\cite{Maldacena:1997re},
namely the correspondence between  gravity in a higher dimensional space and the gauge
theory on its boundary, a large amount of research has been devoted in the investigation
of AdS black holes
\cite{Maldacena:1998bw,Cvetic:1999xp,Chamblin:1999hg,
Caldarelli:1999xj,Horowitz:1999jd,Hemming:2000as,Gibbons:2004ai,Kastor:2009wy}.
On the other hand, it has been widely discussed that if the higher-dimensional
gravitational theory corresponds to modified gravity, then one obtains corrections in the
 gauge theory side in the strong coupling limit \cite{Cai:2001dz}. Finally, the role of
the Maxwell sector has also be found to be important in the aforementioned duality
structure \cite{Chamblin:1999tk,Hawking:1999dp,Gubser:2009qt,Kubiznak:2012wp}.

Spherically symmetric solutions, and in particular AdS black holes, have been studied in
many modified gravity theories
\cite{Aharony:1999ti,Awad:2005ff,Awad:1999xx,Awad:2000ac,Anabalon:2013oea,
Cisterna:2014nua, Babichev:2015rva,
Brihaye:2016lin,Cisterna:2016nwq,Babichev:2017rti,Cvetkovic:2017nkg,Erices:2017izj,
Cisterna:2017jmv, Cisterna:2018mww}, nevertheless
almost all of them remain in the framework of curvature-modified gravity, namely in
modified-gravity formulations which are based on the standard Einstein-Hilbert action. On
the other hand, it was recently realized that one can construct new classes of
gravitational modifications   starting from the torsional formulation of gravity, that is
from the Teleparallel Equivalent of General Relativity (TEGR)
\cite{Cai:2015emx,Linder:2010py,Chen:2010va,
2012GReGr..44.3059M,2012GReGr..44.3059M,Hanafy:2015yya,Zheng:2010am,Bamba:2010wb,
Dent:2011zz,Cai:2011tc,Li:2011rn,
Capozziello:2011hj, Wu:2011kh, Wei:2011aa,Liu:2012fk, Amoros:2013nxa,
Otalora:2013dsa,Bamba:2013jqa,
Li:2013xea,2014EPJP..129..188N,Ong:2013qja,Nashed:2014lva,Kofinas:2014daa,Darabi:2014dla,
Nashed:2015pda, Haro:2014wha,
Hanafy:2014ica,
Guo:2015qbt,Capozziello:2015rda,
Bamba:2016gbu,Malekjani:2016mtm,Farrugia:2016qqe,Nashed:2003ee,
Qi:2017xzl,Sk:2017ucb,
Awad:2017ign, Bahamonde:2017wwk, Awad:2017yod,Karpathopoulos:2017arc,
Hohmann:2017jao,Hohmann:2018rwf,Cai:2018rzd}. Although
black hole solutions in torsional modified gravity, such as $f(T)$ gravity, have been
studied in the literature
\cite{Gonzalez:2011dr,Capozziello:2012zj,Iorio:2012cm,Nashed:2013bfa,Aftergood:2014wla,
Paliathanasis:2014iva,2015IJMPD..2450007N,
2015EPJP..130..124N,
Junior:2015fya,Nashed:2016tbj,Awad:2017sau,Ahmed:2016cuy,Farrugia:2016xcw,Mai:2017riq,
Awad:2017tyz,
Bejarano:2017akj}, the detailed investigation of charged AdS black holes and in
particular in higher dimensions is something that still misses.

In the present work we are interested in investigating rotating charged AdS black holes
in higher-dimensional $f(T)$ gravity.  Such an analysis will be very useful for the
application of  AdS/CFT correspondence in torsional gravity, which may have an
advantage relating to the boundary structure comparing to the curvature formulation
(since the torsion and curvature scalars differ by a boundary term) \cite{Cai:2015emx}.
Furthermore, note
that although curvature-based modified gravity has usually higher-order field equations,
which apart from raising ghost and instabilities issues do not allow for the extraction
of analytic solutions, in the case of $f(T)$ gravity the equations are second-order and
hence analytic solutions can be easily extracted \cite{Cai:2015emx}. This feature
becomes important in relation to the arguments that higher derivatives terms are strongly
constrained in AdS/CFT framework due to causality constraints
\cite{Camanho:2014apa,Papallo:2015rna}. Since $f(T)$ gravity does not include higher
derivatives, it bypasses these constraints, and thus its application to the  AdS/CFT
correspondence has an additional advantage comparing to curvature modified gravity.

The plan of the work is as follows. In Section \ref{S1} we briefly review $f(T)$ gravity
and we extract the equations in the presence of the electrodynamics sector. In Section
\ref{AdsSection} we extract uncharged and charged static AdS solutions, analyzing their
singularity and horizon structure and calculating their energy. In Section \ref{S5} we
extract the
rotating charged AdS solutions in  Maxwell-$f(T)$ gravity. Finally,   Section \ref{S7} is
devoted to the conclusions.

\section{$f(T)$ gravity}
\label{S1}

In this section we present briefly the $f(T)$ gravitational theory. In torsional
formulation of gravity it proves convenient to use the vielbeins fields $b^{i}{_{\mu}}$
(tetrads in four dimensions) as   dynamical variables,
which form an orthonormal basis for the tangent space at each point  of the spacetime.
These are related to the metric through
\begin{equation}\label{q3}
 {\it g_{\mu \nu} :=  \eta_{i j} {b^i}_\mu {b^j}_\nu,}
\end{equation}
 with $\eta_{i j}=(+,-,-,- \cdots)$ being the  $N$-dimensional Minkowskian metric of
the tangent space (Greek indices are used for the coordinate space while Latin
indices for the tangent one).  One  introduces the curvature-less Weitzenb\"{o}ck
connection     $\overset{{\bf{w}}}{\Gamma}^\lambda_{\mu \nu} := {b_i}^\lambda~
\partial_\nu b^{i}_\mu$ \cite{Wr}, and thus the torsion
tensor is defined as
\begin{equation}
\label{torsten}
{T^\alpha}_{\mu \nu} :=
\overset{{\bf{w}}}{\Gamma}^\alpha_{ \nu\mu }-\overset{{\bf{w}}}{\Gamma}^\alpha_{\mu \nu}
={b_i}^\alpha
\left(\partial_\mu{b^i}_\nu-\partial_\nu{b^i}_\mu\right),
\end{equation}
which carries all the information of the gravitational field. Finally, contracting the
torsion tensor we obtain the torsion scalar as
\begin{equation}
\label{Tor_sc}
T\equiv\frac{1}{4}
T^{\rho \mu \nu}
T_{\rho \mu \nu}
+\frac{1}{2}T^{\rho \mu \nu }T_{\nu \mu\rho }
-T_{\rho \mu }^{\ \ \rho }T_{\
\ \ \nu }^{\nu \mu }.
\end{equation}
 When $T$ is used   as the Lagrangian in the action of teleparallel gravity the obtained
theory is the teleparallel equivalent
of general relativity (TEGR), since variation with respect to the vielbeins leads to
exactly the same equations with general relativity.

Inspired by the $f(R)$ extensions of general relativity, one can extend $T$ to $f(T)$,
obtaining  $f(T)$  gravity, determined by the action \cite{Cai:2015emx}:
\begin{equation}\label{q7a}
{\cal L}=\frac{1}{2\kappa}\int |b|f(T)~d^{N}x,
\end{equation}
where $|b|=\sqrt{-g}=\det\left({b^a}_\mu\right)$ is the determinant of the metric and
$\kappa$  is  a dimensional constant defined as $\kappa =2(N-3)\Omega_{N-1} G_N$, where
$G_N$ is the Newtonian
gravitational  constant in $N$-dimensions and
$\Omega_{N-1}$  is the volume of  $(N-1)$-dimensional unit sphere defined as
\begin{equation}
 \Omega_{N-1} = \frac{2\pi^{(N-1)/2}}{\Gamma((N-1)/2)},
 \end{equation}
 with $\Gamma$ the $\Gamma$-function (in the case  $N = 4$ we have
$2(N-3)\Omega_{N-1} = 8 \pi$).

In this work we desire to study the charged AdS black hole solutions in the framework of
$f(T)$ gravity. Hence, in action (\ref{q7a}) we add the Maxwell Lagrangian too.
Therefore, the considered action in this work is
\begin{equation}\label{q7}
{\cal L}=\frac{1}{2\kappa}\int |b|f(T)~d^{N}x+\int |b|{\cal L}_{ em}~d^{N}x,
\end{equation}
where  ${\cal L}_{
em}=-\frac{1}{2}{ F}\wedge ^{\star}{F}$, with $F = dA$ and  $A=A_{\mu}dx^\mu$   the
 electromagnetic potential 1-form \cite{Capozziello:2012zj}.

Variation of action (\ref{q7}) with respect to the vielbeins  leads to
\cite{Cai:2015emx}:
\begin{eqnarray}\label{q8a}
& &\xi^\nu{}_\mu={S_\mu}^{\rho \nu} \partial_{\rho} T
f_{TT}+\left[b^{-1}{b^i}_\mu\partial_\rho\left(b{b_i}^\alpha
{S_\alpha}^{\rho \nu}\right)-{T^\alpha}_{\lambda \mu}{S_\alpha}^{\nu \lambda}\right]f_T
-\frac{f}{4}\delta^\nu_\mu +\frac{1}{2}\kappa{{{\mathfrak{
T}}^{{}^{{}^{^{}{\!\!\!\!\scriptstyle{em}}}}}}}^\nu_\mu \equiv0,
\end{eqnarray}
with $f := f(T)$, \ \   $f_{T}:=\frac{\partial f(T)}{\partial T}$, \ \
$f_{TT}:=\frac{\partial^2
f(T)}{\partial T^2}$,    ${{{\mathfrak
T}^{{}^{{}^{^{}{\!\!\!\!\scriptstyle{em}}}}}}}^\nu_\mu$   the
energy-momentum tensor of the  electromagnetic field defined as
\[
{{{\mathfrak
T}^{{}^{{}^{^{}{\!\!\!\!\scriptstyle{em}}}}}}}^\nu_\mu=F_{\mu \alpha}F^{\nu
\alpha}-\frac{1}{4} \delta_\mu{}^\nu F_{\alpha \beta}F^{\alpha \beta},\]
and  ${S_\mu}^{\nu \alpha}$
the superpotential tensor, which is anti-symmetric in the last
two  indices,
defined as
\begin{equation}\label{q5}
{S_\alpha}^{\mu \nu} :=
\frac{1}{2}\left({{\cal{K}}^{\mu\nu}}_\alpha+\delta^\mu_\alpha{T^{\beta
\nu}}_\beta-\delta^\nu_\alpha{T^{\beta \mu}}_\beta\right),
\end{equation}
  where ${{\cal{K}}^{\mu\nu}}_\alpha$ is the contortion tensor defined as
\begin{equation}\label{q555}
{{\cal{K}}^{\mu \nu}}_\alpha :=
-\frac{1}{2}\left({T^{\mu \nu}}_\alpha-{T^{\nu
\mu}}_\alpha-{T_\alpha}^{\mu \nu}\right).
\end{equation}
Additionally, variation  of   (\ref{q7}) with respect to $A_{\mu}$ gives:
\begin{eqnarray}\label{q8b}
&&\partial_\nu \left( \sqrt{-g} F^{\mu \nu} \right)=0\; .
\end{eqnarray}
The above equations determine Maxwell-$f(T)$ gravity in arbitrary dimensions.

\section{Anti-de-Sitter black hole solutions}
\label{AdsSection}

In this section we extract AdS charged black hole solutions in general dimensions in the
case of $f(T)$ gravity. Using cylindrical  coordinates in $N$ dimensions ($t$, $r$,
$\phi_1$, $\phi_2$,
$\cdots$, $\phi_{n}$, $z_1$, $z_2$ $\cdots$ $z_k$), with  $k=1,2 \cdots$ $N-n-2$, in
which
  $0\leq r< \infty$, $-\infty < t < \infty$, $0\leq \phi_{n}< 2\pi$ and $-\infty <
z_k < \infty$, we
consider the vielbein \cite{Capozziello:2012zj}:
\begin{equation}\label{tetrad}
\hspace{-0.3cm}\begin{tabular}{l}
   $\left({b^{i}}_{\mu}\right)=\left( \sqrt{B(r)}, \; \frac{1}{\sqrt{B_1(r)}}, \; r, \;
r, \; r\;\cdots \right)$,
\end{tabular}
\end{equation}
 which corresponds to the metric
\be
\label{m2}
ds^2=
B(r)dt^2-\frac{1}{B_1(r)}dr^2-r^2\left(\sum_{i=1}^{n}d\phi^2_i+\sum_{k=1}^{N-n-2}
dz_k^2\right),
\ee
where  the functions  $B(r)$ and $B_1(r)$ depend only on the radial
coordinate $r$. 
We mention here that metric (\ref{m2}) is not the most general one since it includes
flat sections and not spherical or hyperbolic ones. This arises from the fact that
considering the general metric in four dimensions one obtains  additionally the
$r-\theta$ field equation which implies that either $f_{TT}=0$ (which is TEGR i.e.
general relativity case), or $T=const.$, or that the $\phi$-section is flat
\cite{Boehmer:2011gw,Iorio:2012cm}. Therefore, in order to obtain general new solutions
we focus on the metric (\ref{m2}).

Substituting the vielbein form (\ref{tetrad}) into the torsion scalar
definition  (\ref{Tor_sc}) we  find
\begin{equation}\label{df}
T=(N-2)\frac{B'B_1}{rB}+(N-2)(N-3)\frac{B_1}{r^2},
\end{equation}
where $B'(r)\equiv\frac{dB(r)}{dr}$ and $B'_1(r)\equiv \displaystyle\frac{dB_1(r)}{dr}$
and from now on  we omit the arguments in $B$,$B_1$,$B'$,$B'_1$. Finally, since the
power-law $f(T)$ form is the one with best agreement with   cosmological data
  \cite{Nesseris:2013jea,Nunes:2016qyp,Basilakos:2018arq}, in the following we focus our
analysis to the
choice
\begin{equation}\label{powellaw}
 f(T)=T+\beta T^2+\gamma T^3-2\Lambda,
\end{equation}
with $\beta$ and $\gamma$ the model parameters and where we have included an explicit
cosmological constant for completeness.

\subsection{Asymptotically static  AdS black holes}\label{S2}

We start our investigation by extracting asymptotically static  $AdS$ black holes in
the case of absent electromagnetic sector, namely considering  ${{{\mathfrak
T}^{{}^{{}^{^{}{\!\!\!\!\scriptstyle{em}}}}}}}^\nu_\mu=0$. In this case, inserting
the vielbein (\ref{tetrad}) into the general field equations (\ref{q8a}),(\ref{q8b}) we
obtain the following non-vanishing components:
\begin{eqnarray}\label{df1}
& &
\!\!\!\!\!\!\!\!\!\!\!\!\!\!\!\!\!\!\!\!
\xi^r{}_r= 2Tf_T+2\Lambda-f=0,\nonumber\\
& &\!\!\!\!\!\!\!\!\!\!\!\!\!\! \!\!\!\!\!\!  \xi^{\phi_1}{}_{\phi_1}=
\xi^{\phi_2}{}_{\phi_2}=\cdots \cdots
=\xi^{\phi_{n}}{}_{\phi_{n}}=\xi^{z_1}{}_{z_1}= \xi^{z_2}{}_{z_2}=\cdots \cdots
=\xi^{z_{N-n-2}}{}_{z_{N-n-2}}\nonumber\\
& &
\!\!\!\!
=   \frac{f_{TT}
[r^2T+(N-2)(N-3)B_1]T'}{(N-2)r}+\frac{f_T}{2r^2{B}^2}\Biggl\{2r^2BB_1B''-r^2B_
1B'^2+2(2N-5)rBB_1B' +r^2BB'B'_1\nonumber\\
& &
\ \ \ \ \ \ \ \ \ \ \ \ \ \ \ \ \ \ \ \ \ \ \ \ \ \ \ \ \ \ \ \ \ \ \ \ \ \ \ \ \ \ \ \ \
\ \ \ \ \ \ \ \ \ \ \ \ \ \ \,
+2(N-3)B^2[2(N-3)B_1+rB'_1]\Biggr\}-f+2\Lambda=0, \nonumber\\
& &
\!\!\!\!\!\!\!\!\!\!\!\!\!\!\!\!\!\!\!\!
\xi^t{}_t=\frac{2(N-2)B_1f_{TT}
T'}{r}+\frac{(N-2)f_T}{r^2B}\Biggl\{2(N-3)BB_1+rB_1B'+rBB'_1\Biggr\}
-f+2\Lambda=0.\nonumber\\
& &
\end{eqnarray}
In the case of the $f(T)$ form (\ref{powellaw}) these equations reduce to
\begin{eqnarray}
\label{df31}
& &
\!\!\!\!\!\!\!\!\!\!\!\!\!
\xi^r{}_r=  T+3\beta T^2+5\gamma T^3+2\Lambda=0,\\
\label{df32}
& &
\!\!\!\!\!\!\!\!\!\!\!\!\!
\xi^{\phi_1}{}_{\phi_1}= \xi^{\phi_2}{}_{\phi_2}=\cdots \cdots
=\xi^{\phi_{n}}{}_{\phi_{n}}=\xi^{z_1}{}_{z_1}= \xi^{z_2}{}_{z_2}=\cdots \cdots
=\xi^{z_{N-n-2}}{}_{z_{N-n-2}}\nonumber\\
  &&
  \ \,
  =\frac{2(\beta+3\gamma T)[r^2T+(N-2)(N-3)B_1]T'}{(N-2)r}+\frac{(1+2\beta T+3\gamma
T^2)}{2r^2{B}
^2}\Biggl\{2r^2BB_1B''-r^2B_1B'^2\nonumber\\
& &
\ \ \ \ \ \ \ \ \ \ \ \
+2(2N-5)rBB_1B'
+r^2BB'B'_1+2(N-3)B^2[2(N-3)B_1+rB'_1]\Biggr\} -T-\beta
T^2-\gamma
T^3+2\Lambda=0,
\\
\label{df33}
& &
\!\!\!\!\!\!\!\!\!\!\!\!\!
\xi^t{}_t= \frac{4(N-2)(\beta+3\gamma T)
B_1T'}{r}+\frac{(1+2\beta T+3\gamma
T^2)(N-2)}{r^2N}\Biggl\{2(N-3)BB_1+rB_1B'+rBB'_1\Biggr\}\nonumber\\
&&
\ \,
-T-\beta T^2-\gamma T^3+2\Lambda=0,
\end{eqnarray}
where $T'\equiv dT(r)/dr$ is calculated through (\ref{df}).

A first observation is that Eq. (\ref{df31}) is a third-order algebraic equation and
hence it implies that
$T=T_0=const.$. Hence, the differential equation  (\ref{df}) for $T=const.$ leads easily
to
the
general solution
\begin{eqnarray}\label{df4g}
& &  B(r)= \Lambda_{eff}r^2-\frac{m}{r^{N-3}},\nonumber\\
&&B_1(r)=B(r)\mathbb{B}\;,
\end{eqnarray}
where $m$ is an integration constant related to the mass parameter, and the function
$\mathbb{B}$ is calculated by inserting (\ref{df4g}) into (\ref{df32}),(\ref{df33}),
giving
 \begin{eqnarray}\label{df4gb}
\mathbb{B}=\frac{T_0\gamma}{(N-1)(N-2)}=const.,
\end{eqnarray}
in the case where we set the explicit cosmological constant $\Lambda$ to zero
(in which case $T_0=\frac{-3\beta\pm\sqrt{9\beta^2-20\gamma}}{10\gamma}$).
In the above expressions the constant $\Lambda_{eff}$
is given by
 \begin{eqnarray}
 \label{Leff}
\Lambda_{eff}=\frac{1}{\gamma},
\end{eqnarray}
 and we can clearly see that it plays the role of an effective cosmological
constant. The important observation here is that we obtain an effective
cosmological constant that arises solely from the $f(T)$ modification,
even if the initial explicit cosmological constant is absent. Hence, interestingly enough,
the structure of the $f(T)$ gravity leads to an effective cosmological
constant and in the case where it is negative the solution is an AdS one. This feature,
namely the induction of an effective cosmological constant due to the $f(T)$ structure,
was indicated to happen in $f(T)$ gravity \cite{Iorio:2012cm,Kofinas:2015hla},
however in the present work we show robustly that it does appear and moreover in general
dimensions.

We mention that the above solution exists only for
$\gamma\neq0$, namely it is a result of the higher-power correction to standard
TEGR, i.e to general relativity, and it reveals the effect of such corrections. In the
case where $\beta=\gamma=0$ and $\Lambda\neq0$ then
$\Lambda_{eff}\propto\Lambda$, that is we recover  standard TEGR with a cosmological
constant, and its Schwarzschild-(A)dS solution.
Lastly, note that although $B(r)$ and $B_1(r)$ differ by a constant, the $g_{tt}$
and $g_{rr}$ components of the metric have  the same  Killing and event horizons. The
solution has a singularity  at $r=0$, while it possesses a horizon at
 $m=\Lambda_{eff} r^{N-1}$.

We continue our analysis for the special choice where
\begin{equation}\label{df5}
\Lambda=\frac{1}{18\beta} \qquad  {\text {and}} \qquad \gamma=\frac{3\beta^2}{5},
\end{equation}
with $\beta\neq0$,
since in this case even for $\Lambda\neq0$ the solution has $B(r)=B_1(r)$.
The investigation of solutions with $g_{tt}
=g^{-1}_{rr}$ has an increased interest in the literature since they have increased
capability in exhibiting the right change of signature in the $t$ and $r$ components for
having an event horizon \cite{Martinez:2005di,Cisterna:2014nua,Bueno:2016lrh}, and
additionally they are the ones that are preferred from Solar System tests
\cite{PhysRevLett.83.3990,Iorio:2012cm}.
In particular,
following the above steps we extract the solution
\begin{eqnarray}\label{df4}
& &  B(r)=\Lambda_{eff}r^2-\frac{m}{r^{N-3}}, \nonumber\\
&&B_1(r)=B(r),
\end{eqnarray}
where $m$ is an integration constant related to the mass parameter and where
 \begin{eqnarray}
 \label{Leff2}
\Lambda_{eff}=-\frac{1}{3(N-1)(N-2)\beta}.
\end{eqnarray}
Similarly to the  previous case   we obtain an effective cosmological constant which
now depends on the parameter $\beta$ and the space dimensionality $N$. In the case where
$\beta>0$ we obtain an AdS solution. The horizon of the solution (\ref{Leff2}) is again
at $m=\Lambda_{eff} r^{N-1}$.

\subsection{New charged AdS black hole solutions}
\label{S3newsol}

Let us now proceed with the analysis of the charged solutions, that is we consider also
the electromagnetic Lagrangian  ${\cal L}_{em}$ in (\ref{q7}), choosing additionally
without loss of generality the vector potential to have the general form $A
= q(r) dt$. Imposing again the
$N$-dimensional vielbeins of  (\ref{tetrad}),  the field equations
(\ref{q8a}),(\ref{q8b}) have the following non-vanishing components:
\begin{eqnarray}
\label{df7}
& &
\!\!\!\!\!\!\!\!\!\!\!\!\!\!\!\!\!\!\!\!\!\!\!\!\!
\xi^r{}_r= 2Tf_T+2\Lambda-f+\frac{2q'^2(r)B_1}{B}=0,\nonumber\\
& & \!\!\!\!\!\!\!\!\!\!\!\!\!\!\!\!\!\!\!\!\!\!\!\!\!
\xi^{\phi_1}{}_{\phi_1}=
\xi^{\phi_2}{}_{\phi_2}=\cdots \cdots
=\xi^{\phi_{n}}{}_{\phi_{n}}=\xi^{z_1}{}_{z_1}= \xi^{z_2}{}_{z_2}=\cdots \cdots
=\xi^{z_{N-n-2}}{}_{z_{N-n-2}}\nonumber\\
&& \!
\!\!\!\!\! \!\!\! = \frac{f_{TT}
[r^2T+(N-2)(N-3)B_1]T'}{(N-2)r}+\frac{f_T}{2r^2{B}^2}\Biggl\{2r^2BB_1B''
-r^2B_1B'^2+4(N-3)^2B^2B_1\nonumber\\
& & \ \ \ \ \ \ \ \ \  \ \ \ \ \ \ \ \ \  \ \ \ \  \ \
+2(2N-5)rBB_1B'
+r^2BB'B'_1+2(N-3)rB^2B'_1\Biggr\} -f+2\Lambda-\frac{2q'^2(r)B_1}{B}=0,
\nonumber\\
& &
\!\!\!\!\!\!\!\!\!\!\!\!\!\!\!\!\!\!\!\!\!\!\!\!\!
\xi^t{}_t= \frac{2(N-2)B_1f_{TT}
T'}{r}+\frac{(N-2)f_T[2(N-3)BB_1+rB_1B'+rBB'_1]}{r^2B}-f+2\Lambda+\frac{2q'^2(r)B_1}{B}=0,
\nonumber\\
& &
\end{eqnarray}
where $q'=\frac{dq(r)}{dr}$.
In the case of the $f(T)$ form (\ref{powellaw}) these equations reduce to
\begin{eqnarray} \label{df7c}
& &
\!\!\!\!\!\!\!\!\!\!\!\!
\xi^r{}_r=  T+3\beta T^2+5\gamma
T^3+2\Lambda+\frac{2q'^2(r)B_1}{B}=0,\nonumber\\
& &
\!\!\!\!\!\!\!\!\!\!\!\!
\xi^{\phi_1}{}_{\phi_1}= \xi^{\phi_2}{}_{\phi_2}=\cdots \cdots
=\xi^{\phi_{n}}{}_{\phi_{n}}=\xi^{z_1}{}_{z_1}= \xi^{z_2}{}_{z_2}=\cdots \cdots
=\xi^{z_{N-n-2}}{}_{z_{N-n-2}}\nonumber\\
&&\ \ =  \frac{2(\beta+3\gamma T)[r^2T+(N-2)(N-3)B_1]T'}{(N-2)r}+\frac{(1+2\beta T+3\gamma
T^2)}{2r^2{B}
^2}\Biggl\{2r^2BB_1B''\!-\!r^2B_1B'^2\!+\!2(2N\!-\!5)rBB_1B'\nonumber\\
& &\ \  \ +r^2BB'B'_1+2(N-3)B^2[2(N-3)B_1+rB'_1]\Biggr\} -T-\beta T^2-\gamma
T^3+2\Lambda-\frac{2q'^2(r)B_
1}{B}=0,\nonumber\\
& &
\!\!\!\!\!\!\!\!\!\!\!\!
\xi^t{}_t= \frac{4(N-2)(\beta+3\gamma T) B_1T'}{r}+\frac{(1+2\beta
T+3\gamma
T^2)(N-2)}{r^2N}\Biggl\{2(N-3)BB_1+rB_1B'+rBB'_1\Biggr\}\nonumber\\
&&
\ \,
-T-\beta T^2-\gamma T^3+2\Lambda+\frac{2q'^2(r)B_1}{B}=0.
\end{eqnarray}
Although the above equations in the case of a general
$\Lambda$ can be solved only numerically, analytical solutions can still
be extracted when $\Lambda$ is given by
 (\ref{df5}). In this case the general
$N$-dimensional solution  takes the form
\begin{eqnarray}\label{df8}
& &
 B(r)=r^2
\Lambda_{eff}-\frac{m}{r^{N-3}}+\frac{15q{}^2(N-3)}{4(N-2)r^{{2(N-3)}}}+\frac{
45(N-3)q{}^3}{16(N-2)(5N-13)\sqrt[3]{r^{2(4N-11)}}}
+\frac{9q{}^4}{8(N-2)(7N-17)\sqrt[3]{r^{2(5N-13)}}},\nonumber\\
&&
B_1(r)=h(r)B(r),
\end{eqnarray}
 with
\begin{eqnarray}
&&h(r)= \left[1+\frac{2q}{3(N-3)\sqrt[3]{r^{{2(N-2)}}}}+\frac
{4q{}
^2}{9(N-3)^2\sqrt[3]{r^{{4(N-2)}}}}
 +\frac{q{}^3}{9(N-3)^3r^{{2(N-2)}}}+\frac{q{}^4}{36(N-3)^4\sqrt[3]{r^{{8(N-2)}}}}
\right]^{-1} ,\nonumber\\
  & &
q(r)=\frac{q}{r^{N-3}}+\frac{q{}^2}{(5N-13)\sqrt[3]{r^{{(5N-13)}}}}+\frac{q{}^3}{
2(N-3)(7N-17)\sqrt[3]{r^{{(7N-17)}}}},
\label{elpot1}
\end{eqnarray}
where
\be \Lambda_{eff}=\frac{81(N-3)^5}{4(N-1)(N-2)q},
\ee
 and with $m$,$q\neq0$ being the constants of integration related to mass and
electric
charge respectively. We stress  that the above solution exists only for
$q\neq0$, since
for $q=0$ we have the solutions of the previous subsection,  thus it arises
from the
structure of the electromagnetic sector.

Let us now discuss on the properties of the above solution.  First of all, inserting
 (\ref{df8}) into  (\ref{tetrad}) and then into (\ref{q3}) we obtain the
metric as
\begin{eqnarray}
\label{metric}
  &&
  \!\!\!\!\!\!\!
  ds{}^2=\Biggl[r^2
\Lambda_{eff}-\frac{m}{r^{N-3}}+\frac{15q^2(N-3)}{4(
N-2)r^{{2(N-3)}}}+\frac{45(N-3)q^3}{16(N-2)(5N-13)\sqrt[3]{r^{2(4N-11)}}}
+\frac{9q^4}{8(N-2)(7N-17)\sqrt[3]{r^{2(5N-13)}}}\Biggr]dt^2\nonumber\\
& &\ \ \
-\Biggl[1+\frac{2q}{3(N-3)\sqrt[3]{r^{{2(N-2)}}}}+\frac{4q^2}{9(N-3)^2\sqrt[3]{r^{{4(N-2)}
}}} +\frac{q^3}{9(N-3)^3r^{{2(N-2)}}}+\frac{q^4}{36(N-3)^4\sqrt[3]{r^{{8(N-2)}}}}\Biggr]
\nonumber\\
&&\ \ \
\times\Biggl[r^2 \Lambda_{eff}\!+\!\frac{m}{r^{N-3}}
\!+\!\frac{15q^2(N-3)}{4(N-2)r^{{2(N-3)}}}\!+\!\frac{45(N-3)q^3}{16(N-2)(5N-13)\sqrt[3]{r^
{
2(4N-11)}}}\!+\!\frac{9q^4}{8(N-2)(7N-17)\sqrt[3]{r^{2(5N-13)}}}\Biggr]^{-1}\!dr^2
\nonumber\\
&&
\ \
-r^2\left(\sum_{i=1}^{n}d\phi^2_i+\sum_{k=1}^{N-n-2}dz_k^2\right).
\end{eqnarray}
As we observe, in this case the solution is more complicated, however it is still
asymptotically AdS or dS according to the sign of $q$. We mention the interesting
feature that the effective cosmological constant is a result of the  electric
charge. Nevertheless, note that both
$\beta$ and $\gamma$ are important for the solution structure, and hence this solution
does not have a TEGR, i.e. general relativity, limit, nor an uncharged one. Moreover,
this solution subclass is
also outside the ones obtained in \cite{Capozziello:2012zj,Awad:2017tyz}, due to the use
of more general $f(T)$ forms in the present work.
Therefore, solution  (\ref{df8}) corresponds to a new charged AdS black hole in power-law
$f(T)$ gravity.

Additionally, we mention here that apart from the difference in the metric part, the above
solution has also a difference in the  charge structure too, comparing to those obtained
in \cite{Gonzalez:2011dr,Capozziello:2012zj,Awad:2017tyz}, since the potential $q(r)$
depends on a monopole and higher-order electromagnetic potential. This electromagnetic
potential will have a vanishing value only when the constant $q=0$, which is not allowed
in the above solution, and thus this implies that within the framework of $f(T)$ gravity
we cannot find a charged solution with monopole only.

We now proceed to the investigation of the singularity structure of the solution, by
calculating curvature and torsion invariants. The curvature scalars are calculated from
the
metric (\ref{metric}) while the torsion scalar is calculated through the vielbeins
(\ref{tetrad}), or
straightaway from (\ref{df}). Additionally, observing the solution  (\ref{df}) we deduce
that it is adequate to focus our analysis close to the roots of the function $h(r)$.
Calculating  the Ricci scalar,  the Ricci tensor square, and the Kretschmann
scalar, we respectively find:
\begin{eqnarray}
R=
F_1(r)\,\left(\frac{1}{\sqrt[3]{r^{2(N-2)}}}\right),
 \qquad R^{\mu \nu}R_{\mu \nu}=
F_2(r)\,\left(\frac{1}{\sqrt[3]{r^{4(
N-2)}}}\right),
\qquad
K\equiv  R^{\mu \nu \lambda \rho}R_{\mu \nu \lambda \rho}=
F_3(r)\,\left(\frac{1}{\sqrt[3]{r^{4( N-2)}}}\right),
\end{eqnarray}
while calculating the torsion scalar we respectively obtain:
\begin{eqnarray}
T(r)=\frac{12\sqrt[3]{r^{2(N-2)}}+\beta[
3(N-3)]^5(N-2)}{36\left|\beta\right| \sqrt[3]{r^{2(N-2)}}},
\end{eqnarray}
where  $F_i(r)$ are polynomial functions of $r$.
The above invariants first of all show the  singularity at $r=0$. Close to $r=0$ the
behavior of these invariants are
given by $(K,R_{\mu
\nu}R^{\mu \nu}) \sim \sqrt[3]{r^{-4(N-2)}}$  and  $(R,T)\sim  \sqrt[3]{r^{-2(N-2)}}$, in
contrast to the  solutions of the  Einstein-Maxwell theory in
both general relativity and TEGR formulations which have $(K ,R_{\mu \nu}R^{\mu \nu})\sim
r^{-2N}$ and $(R,T) \sim  r^{-N}$. This shows clearly
that the
singularity of our charged solution is softer than the one obtained in GR and TEGR for
the charged
case. Finally, notice that although in the solution the
$g_{tt}$ and $g_{rr}$ components of the metric are different, they have the same
Killing  and event horizons.

\begin{figure}[ht]
\centering
\includegraphics[scale=0.4]{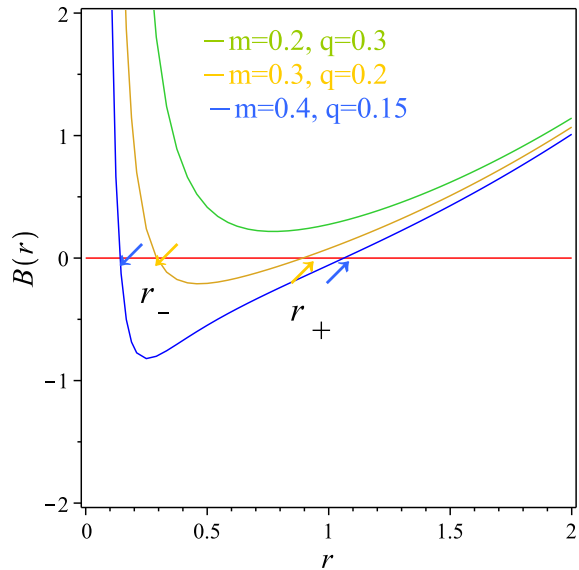}
\caption{{\it{The metric function $B(r)$ of solution (\ref{df8}) of Maxwell-$f(T)$
gravity in four dimensions, for various values of the mass parameter $m$ and the electric
charge $q$ in units
where $\kappa=1$. $r_-$ denotes the black hole inner Cauchy horizon   and $r_+$ the black
hole outer event  horizon.}}
 }
\label{Fig:1}
\end{figure}
In a similar way we can investigate the horizons of Eq. (\ref{df8}), which can
alternatively be calculated examining the roots of $B(r) = 0$.
In the case of four dimensions in Fig. \ref{Fig:1} we depict  $B(r)$ of   solution
(\ref{df8}), for various values of the model parameters. From this graph we can see the
two roots of $B(r)$ that define the black hole inner Cauchy horizon $r_-$ and the black
hole outer event  horizon $r_+$ \cite{Brecher:2004gn}.
As we observe, as $q$ increases and $m$ decreases, and in particular for $q>m$, we enter
in a parameter region where there is no horizon, and thus the central singularity is a
naked singularity. This is an interesting result of Maxwell-$f(T)$ gravity (we mention
that we see the issue from the mathematical point of view and we do not examine whether
such a solution can indeed be formed physically through gravitational collapse), which
does not appear in the absence of the electromagnetic sector (indications of this feature
had also been discussed in
\cite{Gonzalez:2011dr,Capozziello:2012zj}). Moreover, note   that for suitable $m$ and
$q$ the two horizons coincide and become degenerate, namely we obtain  $r_-=r_+\equiv
r_{dg}$.

\begin{figure}[ht]
\centering
\includegraphics[scale=0.4]{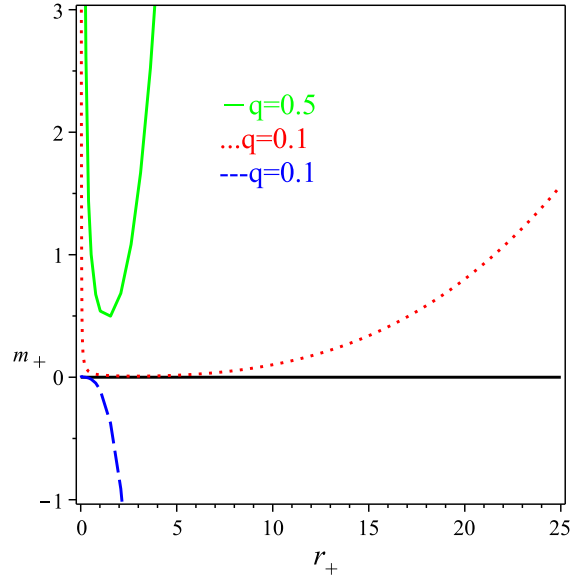}
\caption{\it{{The value $m_+$ of the parameter $m$  that corresponds to the horizon
$r_+$, of solution (\ref{df8}) of Maxwell-$f(T)$
gravity  in four dimensions, for various values of the electric charge $q$  in units
where $\kappa=1$.  The horizontal line at $m_+=0$  is drawn for convenience. }}
 }
\label{Fig:2}
\end{figure}

Finally, in order to present the above features in a different way,  in Fig. \ref{Fig:2}
we depict the value $m_+$ of the parameter $m$  that corresponds to the horizon $r_+$,
which is obtained setting $B(r_+)=0$, namely
\begin{equation} \label{hor1}
{m_+}=r_+^{(N-3)}\left(r_+^2\Lambda_{eff}+\frac{15q{}^2(N-3)}{4(N-2)r_+^{{2(N-3)}}}
+\frac{
45(N-3)q{}^3}{16(N-2)(5N-13)\sqrt[3]{r_+^{2(4N-11)}}}
+\frac{9q{}^4}{8(N-2)(7N-17)\sqrt[3]{r_+^{2(5N-13)}}}\right).\nonumber\\
\end{equation}

\subsection{Energy of the AdS black holes}\label{S3energy}

In this subsection we discuss the energy issues of the obtained solutions. In general
$f(T)$ gravity and for general geometry, equation  (\ref{q8a}) can be rewritten   as
\be
\label{energ1}
\partial_\beta \Biggl[b{S}^{a \rho \beta} f_T\Biggr]=\kappa b
{b^a}_\beta \Biggl[t^{\rho \beta}+{{{\cal
T}^{{}^{{}^{^{}{\!\!\!\!\scriptstyle{em}}}}}}}^{\rho \beta}\Biggr],
\ee
where $t^{\alpha \beta}$ is   defined
as  \cite{Maluf:2002zc,Maluf:1995re}
\be
t^{\alpha
\beta}:=\frac{1}{\kappa}\Biggl[4f_T {S^\mu}^{\beta
\lambda}{T_{\mu \lambda}}^{\alpha}-g^{\alpha \beta} f\Biggr].
\ee
Due to the fact that the
superpotential ${S}^{a \nu \lambda}$ is antisymmetric in the last two indices,
  we
have  $
\partial_\alpha \partial_\beta \left[b{S}^{a \alpha \beta} f_T\right]=0$,
which using (\ref{energ1}) gives
 $ \partial_\beta[b(t^{a \beta}+{{{\cal
T}^{{}^{{}^{^{}{\!\!\!\!\scriptstyle{em}}}}}}}^{a
\beta})]=0$,
which leads to
  \be
  \label{2}
\frac{d}{dt}\int_V d^{(N-1)}x \ b \ {b^a}_\alpha \left(t^{0 \alpha}+{{{\cal
T}^{{}^{{}^{^{}{\!\!\!\!\scriptstyle{em}}}}}}}^{0
\alpha}\right)+ \oint_\Sigma \left[b \ {b^a}_\alpha \ \left(t^{j
\alpha}+{{{\cal T}^{{}^{{}^{^{}{\!\!\!\!\scriptstyle{em}}}}}}}^{j
\alpha}\right)\right]=0.
\ee
Hence, we can now interpret    Eq. (\ref{2})  as  the  conservation law of  the
energy-momentum tensor ${{{\cal
T}^{{}^{{}^{^{}{\!\!\!\!\scriptstyle{em}}}}}}}^{\lambda
\mu}$  and       $t^{\lambda \mu}$. Therefore,
 $t^{\lambda \mu}$ can
be interpreted as  the energy-momentum tensor of the gravitational field in
  $f(T)$    theory \cite{Maluf:2002zc,Maluf:1995re}.
  Hence,
the   energy-momentum of $f(T)$  gravity
in  $(N-1$)-dimensional volume $V$ writes as
\be
\label{3}
P^a=\int_V d^{(N-1)}x
\ b \ {b^a}_\alpha \left(t^{0 \alpha}+{{{\cal
T}^{{}^{{}^{^{}{\!\!\!\!\scriptstyle{em}}}}}}}^{0
\alpha}\right)=\frac{1}{\kappa}\int_V d^{(N-1)}x  \partial_\beta\left[b{S}^{a 0
\beta} f_T\right].
\ee
Note that the above expressions recover the known results of TEGR in the case $f(T)=T$.

Let us now apply the above general analysis in the specific class of power-law $f(T)$
gravity and for the AdS black hole solutions obtained above, namely for
expressions (\ref{df4}) and (\ref{df8}). Inserting   solution (\ref{df4})  into the above
general expressions gives
\begin{equation}  \label{7}
 P^0=E=
 \frac{(N-2)[1+2(N-1)(N-2)\Lambda_{eff}\beta+6(N-1)(5N^2-33N+58)\gamma
\Lambda_{eff}{}^2]m }{
4(N-3)G_N}+{\mathcal{O}}\Biggl(\frac{1}{r}\Biggr),
\end{equation}
where we have replaced $\kappa$ by  $\kappa =2(N-3)\Omega_{N-1} G_N$. Similarly,
inserting  (\ref{df8}) into the  general energy  expressions gives
\begin{equation}  \label{9}
 E= \frac{(N-2)[1+2(N-1)(N-2)\Lambda_{eff}\beta+6(N-1)(5N^2-33N+58)\gamma
\Lambda_{eff}{}
^2]m }{4(N-3)G_N}
+{\mathcal{O}}\Biggl(\frac{1}{r^{4/3}}\Biggr).
\end{equation}
As we observe the $f(T)$ modification has an effect on the mass term of the energy of
standard TEGR  \cite{Maluf:2002zc,Maluf:1995re}, while the charge term does not appear up
to ${\mathcal{O}}\Bigg(\frac{1}{r}\Bigg)$. The charge term will contribute to the
calculation of energy
starting from ${\mathcal{O}}\Bigg(\frac{1}{r^{4/3}}\Bigg)$, in contrast to
Reissner-Nordstr\"om
spacetime. This difference comes from the contribution of the function $h(r)$ given in
relation (\ref{df8}).

 \section{Rotating black holes in Maxwell-$f(T)$ gravity }\label{S5}

 We close this work by  deriving rotating solutions that satisfy the field
equations of power-law $f(T)$ gravity. In order to do so, we will be based on the static
solutions extracted above. In particular, we apply the following
  transformations with $n$
rotation parameters:
\be \label{t1}
\bar{\phi}_{i} =-\aleph~ {\phi_{i}}+\frac{ a_i}{l^2}~t,\qquad \qquad \qquad
\bar{t}=
\aleph~ t-\sum\limits_{i=1}^{n}a_i~ \phi_i,
\ee
with  $a_i$  the   rotation parameters (their number is
$n= \lfloor(N - 1)/2\rfloor$ where  $\lfloor ... \rfloor$ marks the
integer part), and where the parameter $l$ is related to the parameter
$\Lambda_{eff}$ of the static solution
through
\begin{eqnarray}
l
=-\frac{(N-2)(N-1)}{2 \Lambda_{eff}}.
\end{eqnarray}
Additionally,
$\aleph$ is defined as
\[\aleph:=\sqrt{1-\sum\limits_{j=1}^{{n}}\frac{a_j{}^2}{l^2}}.\]
Applying the transformation (\ref{t1})
to the vielbein (\ref{tetrad}) we obtain
\begin{eqnarray}\label{tetrad1}
\nonumber \left({e^{i}}_{\mu}\right)=\left(
  \begin{array}{cccccccccccccc}
    \aleph\sqrt{B(r)} & 0 &  -a_1\sqrt{B(r)}&-a_2\sqrt{B(r)}\cdots &
-a_{n}\sqrt{B(r)}&0&0&\cdots&0
\\[5pt]
    0&\frac{1}{\sqrt{B_1(r)}} &0 &0\cdots &0&0&0&\cdots & 0\\[5pt]
          \frac{a_1r}{l^2} &0 &-\aleph r&0  \cdots &0&0&0&\cdots & 0\\[5pt]
        \frac{a_2r}{l^2} &0 &0  &-\aleph r\cdots & 0&0&0&\cdots & 0\\[5pt]
        \vdots & \vdots  &\vdots&\vdots&\vdots &\vdots&\vdots& \cdots & \vdots \\[5pt]
  \frac{ a_nr}{l^2}  &  0 &0&0 \cdots & -\aleph r&0&0&\cdots & 0 \\[5pt]
   0 &  0  &0&0 \cdots &0&r&0&\cdots & 0\\[5pt]
     0 &  0  &0&0 \cdots &0&0&r&\cdots & 0\\[5pt]
       0 &  0 &0&0 \cdots &0&0&0&\cdots & r\\
  \end{array}
\right),&\\
\end{eqnarray}
   where $B(r)$ and $B_1(r)$ are given by the previously extracted static
solution (\ref{df8}).  The vielbein (\ref{tetrad1}) coincides with (\ref{tetrad})
when
the
rotation parameters $a_i=0$. Hence, for the electromagnetic potential
(\ref{elpot1}) we obtain the form
\be
\label{Rotpot}
\bar{q}(r)=-q(r)\left[\sum\limits_{j=1}^{n} a_j d\phi'_j-\aleph
dt'\right].
\ee
Note here that although the transformation (\ref{t1}) leaves the local
properties of spacetime unaltered, it does change them globally as has been shown
in \cite{Lemos:1994xp}, since it mixes compact and noncompact coordinates. Thus,
the vielbein (\ref{tetrad}) and (\ref{tetrad1}) can be locally mapped into each
other but not globally \cite{Lemos:1994xp,Awad:2002cz}. 

The metric that corresponds to  the vielbein (\ref{tetrad1}) is written as
\ba
\label{m1}
    ds^2=-B(r)\left[\aleph d{t'}  -\sum\limits_{i=1}^{n}  a_{i}d{\phi}'
\right]^2+\frac{dr^2}{B_1(
r)}+\frac{r^2}{l^2}\sum\limits_{i=1 }^{n}\left[a_{i}d{t}'-\aleph l^2 d{\phi}'_i\right]^2+
r^2 dz_k^2-\frac{r^2}{l^2}\sum\limits_{i<j
}^{n}\left(a_{i}d{\phi}'_j-a_{j}d{\phi}'_i\right)^2,
\ea
where $0\leq r< \infty$, $-\infty < t < \infty$, $0 \leq \phi_{i}< 2\pi$, $i=1,2 \cdots
n$ and $-\infty < z_k < \infty$, and where  $d z_k^2$ is
the Euclidean metric on $(N-n-2)$ dimensions with $k = 1,,2\cdots N-3$. Note that
the static configuration (\ref{m2}) can be
recovered as a special case of the above general metric when the rotation parameters
$a_j$ are chosen to be vanished, while by inverting the coordinate
transformations
(\ref{t1}) we get back the static spacetime (\ref{metric}).

  We mention that  the
transformation (\ref{t1}) can be carried out locally but not
globally, since  the first
Betti
number of the manifold is one due to the fact that closed curves encircling the horizon
cannot be shrunk
to zero \cite{PhysRevD.26.1281,Bonnor_1980}. In both  static
 (\ref{metric}) and stationary  (\ref{m1}) spacetimes   there is a timelike
Killing
field $\xi= \frac{\partial}{\partial t}$. In the static spacetime this corresponds to an
exact one-form $V$ inverse
to $\xi$ (i.e. $\bar{V}_\mu\equiv \frac{\xi_\mu}{\| \xi \|^2}$) given then by $\bar{V} =
dt$ (see \cite{Bonnor_1980} for details), while in the stationary
spacetime the corresponding one-form is $\bar{V} = dt +a_i d\phi_i$ which is a closed
one-form but not exact. De Rham's cohomology theorems then state that since the first
Betti number of the manifolds is one, there are global diffeomorphisms which map
the $\xi$ of the two manifolds, however there is no such global diffeomorphisms mapping
$\bar{ V}$
and $V$. Hence, since the metric maps vectors into one-forms, it is implied  that metrics
(\ref{metric}) and (\ref{m1})   can be locally mapped into each other but not globally,
and thus they are distinct.

 Additionally, note that the line-element (\ref{m1}) is
created when the Minkowskian metric in (\ref{q3})
takes in cylindrical  coordinates the  form
\begin{eqnarray}
\label{min}
\nonumber \left(\eta_{ij}\right)=\left(
  \begin{array}{cccccccccccccc}
    -1 & 0 & 0&0&0&0&0&0&0&\cdots&0 \\[5pt]
    0&1 &0 &0 &0&0&0&0&0&\cdots & 0\\[5pt]
         0 &0 &1+\frac{a_n{}^2}{l^2\aleph^2} &-\frac{a_na_1}{l^2\aleph^2}
&-\frac{a_na_
2}{l^2\aleph^2}&\cdots &-\frac{a_na_{n-1}}{l^2\aleph^2}&0&0&\cdots &0\\[5pt]
        0 &0 &-\frac{a_na_1}{l^2\aleph^2}
&1+\frac{a_{n-1}{}^2}{l^2\aleph^2} & -\frac{
a_{n-1}a_1}{l^2\aleph^2}&\cdots&-\frac{a_{n-1}a_{n-2}}{l^2\aleph^2}
&0&0&\cdots & 0\\[
5pt]
        \vdots & \vdots  &\vdots&\vdots&\vdots &\vdots&\vdots& \cdots & \vdots \\[5pt]
  0 &0 &-\frac{a_na_{n-1}}{l^2\aleph^2} &-\frac{a_{n-1}a_{n-2}}{l^2\aleph^2}
&-\frac{a_{
n-2}a_{n-3}}{l^2\aleph^2}&\cdots &1+\frac{a_1{}^2}{l^2\aleph^2}&0&0&\cdots
&0\\[5pt]
   0 &  0  &0&0  &0&\cdots&0&1&0&\cdots & 0\\[5pt]
     0 &  0  &0&0 &0&\cdots&0&0&1&\cdots & 0\\[5pt]
       0 &  0 &0&0 &0&\cdots&0&0&0&\cdots & 1\\
 \end{array}
\right).&\\
\end{eqnarray}
It is of interest to note that the torsion components of the above Minkowski metric
are vanishing.

In summary, we have managed to extract the rotating charged AdS black hole solution in
power-law $f(T)$ gravity. This is a novel solution and one of the main results of the
present paper. Concerning the singularity properties, as we observe from
the structure of  (\ref{m1}), this will be the same with the static solution of
(\ref{metric}). Hence, all the results and the discussion in  subsection
\ref{S3newsol} are valid for the rotating solutions above too. Thus,  at $r=0$ we obtain a
singularity, and close to $r=0$ the
behavior of the invariants is $(K,R_{\mu
\nu}R^{\mu \nu}) \sim \sqrt[3]{r^{-4(N-2)}}$ and  $(R,T)\sim  \sqrt[3]{r^{-2(N-2)}}$, in
contrast to the charged solutions of general relativity and TEGR theories.
Additionally, the horizon structure is
qualitatively similar to the one discussed in   the end of subsection
\ref{S3newsol}, and we do obtain the appearance of a
naked singularity for suitable large $q$. Finally, concerning the energy of the rotating
charged
AdS black
hole (\ref{m1}), following the procedure of subsection \ref{S3energy} it is calculated as
\begin{equation}  \label{9}
 E= \frac{(N-2)[1+2(N-1)(N-2)\Lambda_{eff}\beta+6(N-1)(5N^2-33N+58)\gamma
\Lambda_{eff}{}
^2][\Lambda_{eff}{}^2 a_j\sum\limits_{j=1}^{{n}}a_j+3\aleph^2]m }{12(N-3)G_N}
+{\mathcal{O}}\Biggl(\frac{1}{r^{4/3}}\Biggr).
\end{equation}

 \section{Conclusions}
 \label{S7}

After the formulation of the AdS/CFT correspondence there is an increasing interest in
the extraction and study of (higher-dimensional) Anti-de-Sitter black hole solutions.
Apart from standard gravity, the investigation extends to various gravitational
modifications as well as in the case where the Maxwell sector is also present.
Nevertheless, almost all of the works remain in the framework of curvature-modified
gravity. Thus, in the present manuscript we investigated static and rotating, uncharged
and charged, AdS black holes in higher-dimensional $f(T)$ gravity, focusing on the
power-law
ansatz which is the most viable according to observations.

In the case where the electromagnetic sector is absent we extracted AdS static solutions,
which are characterized by an effective cosmological constant that depends on the
$f(T)$ modification. In the case where we switch on the Maxwell sector, we
analytically obtained charged static solutions, which are asymptotically AdS,
characterized by an effective cosmological constant that depends on the parameters of the
$f(T)$ modification, as well as on the electric charge. Hence, these solution subclasses
do not have a TEGR, i.e. general relativity limit, nor an uncharged one, and
they correspond to  new charged AdS black holes in power-law $f(T)$ gravity, where their
features arise solely from the torsional modification alongside the Maxwell sector
incorporation. Finally, we showed that in the extracted solutions the potential $q(r)$
depends on a monopole and higher-order electromagnetic potential, and thus
within the framework of $f(T)$ gravity we cannot find a charged solution with monopole
only.

As a next step we examined the singularity structure of the solutions, calculating the
values of various curvature and torsion invariants, showing that they do possess the
singularity at $r=0$, which however  is softer comparing to the standard general
relativity case due to the  $f(T)$ structure. Additionally, we investigated the horizons
of the solutions, showing that the solutions possess an event horizon as well  as a
cosmological  horizon. Nevertheless, for suitably large electric charge and small mass we
obtain the  the appearance of a naked
singularity. Finally, we calculated the energy of
the obtained solutions, showing that the $f(T)$ modification affects the mass term.

Based on the analysis of static solutions, through suitable transformations we were able
to extract rotating AdS solutions of Maxwell-$f(T)$ gravity. Similarly to the static
case, the effective cosmological constant arises from the $f(T)$ modification and the the
Maxwell sector. Additionally, the singularity  and horizon properties are the same with
the uncharged case. These solutions also do not have a TEGR or an uncharged limit, and
they correspond to a novel class.

 The extraction of AdS black holes in $f(T)$ gravity may be very helpful towards the
investigation of AdS/CFT correspondence in torsional framework. This direction is
expected to have advantages comparing to the standard curvature-based formulation, due to
the known relations between curvature and torsion invariants through boundary terms.

\subsection*{Acknowledgments}
 G.N. would like to thank   A. Awad and  W. El Hanafy for useful discussions.  This
work is partially supported by the Egyptian Ministry of Scientific Research under project
No. 24-2-12.
This article is based upon work from COST Action CA15117 ``Cosmology and Astrophysics
Network for Theoretical Advances and Training Actions'' (CANTATA), supported by COST
(European Cooperation in Science and Technology).

%

\end{document}